# Exploring Scientists' Working Timetable: A Global Survey

Xianwen Wang[1,2]*, Lian Peng[1†], Chunbo Zhang[1†], Shenmeng Xu[1†], Zhi Wang[1], Chuanli Wang[1], Xianbing Wang[3]

[1]WISE Lab, Dalian University of Technology, Dalian 116085, China.

[2]DUT- Drexel Joint Institute for the Study of Knowledge Visualization and Scientific Discovery, Dalian University of Technology, Dalian 116085, China.

[3]School of Control Science and Engineering, Dalian University of Technology, Dalian 116085, China.

† These authors contributed equally as second authors.

* Corresponding author. Tel.: +86 411 847 072 92-23; fax: +86 411 847 060 82.

Email address: xianwenwang@dlut.edu.cn

**Abstract:** In our previous study (Wang et al., 2012), we analyzed scientists' working timetable of 3 countries, using realtime downloading data of scientific literatures. In this paper, we make a through analysis about global scientists' working habits. Top 30 countries/territories from Europe, Asia, Australia, North America, Latin America and Africa are selected as representatives and analyzed in detail. Regional differences for scientists' working habits exists in different countries. Besides different working cultures, social factors could affect scientists' research activities and working patterns. Nevertheless, a common conclusion is that scientists today are often working overtime. Although scientists may feel engaged and fulfilled about their hard working, working too much still warns us to reconsider the work-life balance.

# 1. Introduction

Scientists are living and breathing their work due to the strong competition and various pressures (Hagstrom, 1974; Hermanowicz, 2003; Hermanowicz J, 2009). In the science community, the participants need to have a high level of work devotion (Fox & Stephan, 2001;Blair-Loy, 2004). They need not only to accomplish their duty and work, but also to continue learning to hold the lead in their research fields in this day-to-day environment.

On the other hand, the rapid growth of scientific literature makes scientists' time more compressed. To keep track of all new progress of research, scientists have to spend more time on searching and reading papers.

The work–life balance contributes to more attrition among scientists (National Research Council, 1998; Jacobs & Winslow, 2004; Post et al., 2009; Fox, Fonseca, & Bao, 2011). However, because it is really difficult to know scientists' working habits, previous studies are scarce. Except questionnaire survey (Petkova & Boyadjieva, 1994; Frone, 2000) or follow-up research (Ledford, 2011), there are hardly good methods to investigate on this topic (Schiermeier, 2012). In our previous initial study (Wang et al., 2012), by using the function of the Springer Realtime platform (http://realtime.springer.com/map), we depict and record the location (as determined by IP-to-city matching) of downloads in real time once one article, book chapter, protocol, or image is downloaded from Springer. In the previous study, we focused on 3 main countries, which are the United State, Germany and Mainland China, and tried to explore scientists' working timetable.

On this basis, we conduct further research in a world wide scope. Representative countries/territories from America, Africa, Australia, Asia, and Europe are included in this study.

## 2. Data and methods

Realtime downloading data are recorded and collected 24/7 from realtime.springer.com. The number of records we have collected in this study is as much as approximately 1.87 million. It is necessary to apply relation database management system. So a SQL database is designed in Microsoft SQL Server specifically for this study. All the collected data are stored, processed in the local database. For the statistical analysis, queries are executed directly from SQL Server. i.e. querying downloads number per minute, per 10 minute and per half hour.

**2.1 Data collecting**

Realtime.springer.com is an analytics platform launched by Springer Verlag since December 2010. It provides four kinds of visualizations of the usage data that is generated worldwide by Springer's literatures.

Realtime.springer.com aggregates the data on downloads of Springer journal articles, images, book chapters, and protocols in real time from scientists all over the world, and displays them in a variety of interactive visualizations. It receives feeds from: SpringerLink, SpringerImages, and SpringerProtocols. For SpringerLink and SpringerProtocols, when a user is downloading the PDF or HTML version of an article, chapter, or protocol, the platform of realtime.springer.com would get a message from the springer sites. For images, a "download" is counted when an Image Details page is shown with the full version (http://realtime.springer.com/about).Visualizations include a constantly updating Keywords Tag Cloud page showing the most frequent downloaded keywords, a Map page showing where the downloads are coming from, a Feed page providing the latest downloaded items, and an Icons page showing the source of downloads. In this paper, we use the same dataset harvested in our previous research (Wang et al., 2012).

In every single minute, there maybe tens of downloads from somewhere over the world. Accordingly, we record the raw data in 3 groups, in order to complement each other in the data processing. In this way, we ensure the greatest degree of data integrity.

**2.2 Data processing**

Our initial data are recorded in Greenwich Mean Time (GMT). Nevertheless, in order to better observe and analyze the working timetable of scientists in each region, we need to convert the GMT time to local time. Unlike the analysis of only top 3 countries in the previous paper (Wang et al., 2012), here we need to convert the time of every region all over the world.

Consequently, we need to identify the time zone of each city first. According to the latitude and longitude data, we get the time zones of all the 5931 cities which have downloads during our study period. Then, we calculate the local time from the GMT based on the corresponding time zone of the city.

Cities in some countries, including most Asian countries, most European countries, all African countries, etc., use only one single time zone. In this case, time conversion is relatively easy, for we just need to convert the GMT to a specific city's local time according to the uniform time zone of that country. Nevertheless, some countries span several time zones, so we have to get the time zone of each city separately. In addition, for some other countries or territories, the time transformation is even more complicated. For example, the cities located in India, Iran, Newfoundland, Afghanistan, Venezuela, Burma, the Marquesas, as well as parts of Australia use half-hour deviations from standard time; and few nations such as Nepal, some individual regions such as the Chatham Islands, use quarter-hour deviations. The time conversion is shown in Table 1. For the few data with the missing field of city names, we identify them according to the latitude and longitude.

**Table 1** Time conversion of worldwide cities

| City | Country/ territories | Latitude | Longitude | GMT time | offset | Local time |
|---|---|---|---|---|---|---|
| Madrid | Spain | 40.4 | -3.683 | 4/9/2012 0:17 | 2 | 4/9/2012 2:17 |
| Berlin | Germany | 52.517 | 13.4 | 4/9/2012 0:16 | 2 | 4/9/2012 2:16 |
| London | UK | 51.517 | -0.105 | 4/9/2012 0:32 | 1 | 4/9/2012 1:32 |
| Zurich | Switzerland | 47.367 | 8.55 | 4/9/2012 0:17 | 2 | 4/9/2012 2:17 |
| Stanford | USA | 37.4162 | -122.172 | 4/9/2012 0:56 | -7 | 4/8/2012 17:56 |
| Leuven | Belgium | 50.883 | 4.7 | 4/9/2012 0:49 | 2 | 4/9/2012 2:49 |
| Seoul | South Korea | 37.567 | 127 | 4/9/2012 0:37 | 9 | 4/9/2012 9:37 |
| - | South Korea | 37.567 | 127 | 4/9/2012 1:00 | 9 | 4/9/2012 10:00 |
| Taipei | Taiwan | 25.017 | 121.45 | 4/9/2012 1:00 | 8 | 4/9/2012 9:00 |
| Beijing | China | 39.9 | 116.413 | 4/9/2012 0:24 | 8 | 4/9/2012 8:24 |
| Leiden | Netherlands | 52.15 | 4.5 | 4/9/2012 0:24 | 2 | 4/9/2012 2:24 |
| Tokyo | Japan | 35.7 | 139.767 | 4/9/2012 0:24 | 9 | 4/9/2012 9:24 |
| Adelaide | Australia | -34.93 | 138.60 | 4/9/20120:18 | 9.5 | 4/9/20129:48 |

When processing the data, we find some abnormal downloads from certain countries. Fig. 1(a) displays the normal download curves of Beijing, China. The curves show a rhythm mostly in step with its country, and there are no distinct differences between each weekday curve and weekend curve. Fig. 1(b) shows the download curves of Tianjin, another city in China. During the period of 23:00, April 10 to 9:00, April 11, which is generally considered as sleeping time, the average downloads per 10 minutes reach as much as 122.8, decupling the downloads in the daytime. Similar abnormal downloads are detected in other regions. The downloads from Los Angles (USA) on April 21 and 22, and downloads of Douglas (UK) on April 14, 15 and 22 are also abnormal. For these abnormal data, which only account a low proportion of the total data, we replace the data with the city's normal data in the same time period on similar weekdays. For example, we replace the data of Los Angles on April 21 and 22 with the data on April 14 and 15. The percentage of replaced data in all is not greater than

1.75%.

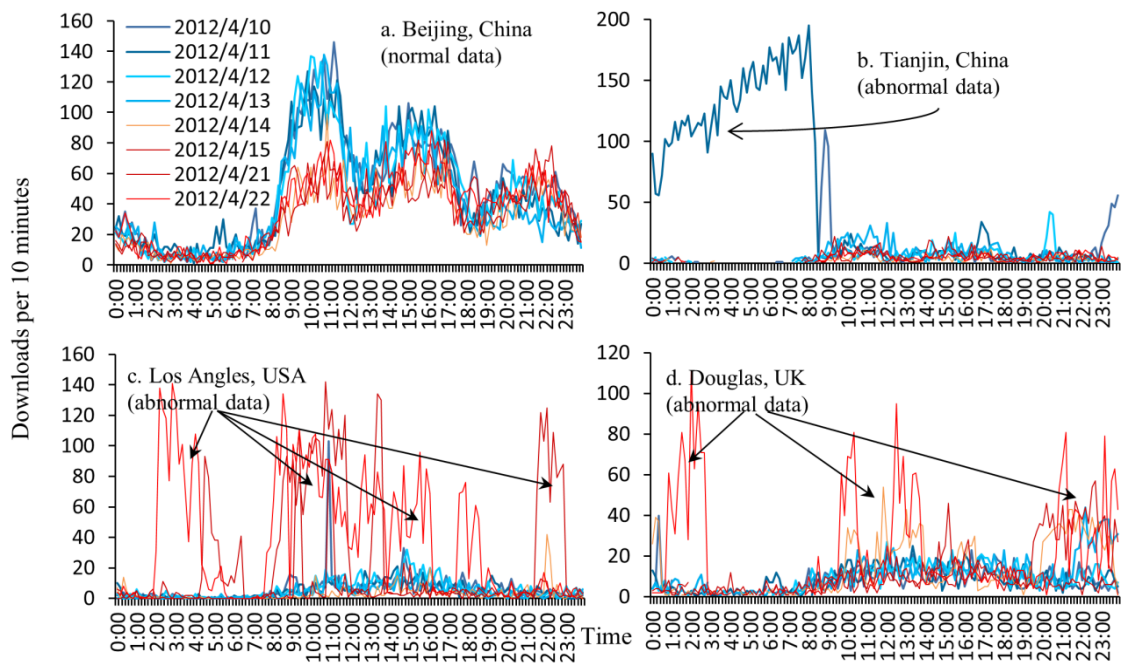

**Fig. 1** Processing of abnormal data

## 3. Results

**3.1 The geography of total downloads**

Table 2 lists the total number of downloads of the top 30 countries/territories during our study period (April 10 –15, April 21 –22 in 2012). The United States has downloaded 392,985 items from Springer, ranking the first; then it follows Germany, who has downloaded 179,660 items; Mainland China has 151489 download times and ranks the third. The top 10 countries/territories also include the United Kingdom, Canada, Australia, Japan, India, Iran, and France. Among the top 30 countries/territories listed in Table 2, 13 are from Europe, and 10 are from Asia. In addition, North America/ Australia/ Latin America have 2 countries respectively, while Africa has only 1 country.

**Table 2** Downloads of top 30 countries/territories (April 10 – April 15, April 21 – April 22, 2012, GMT)

| rank | country/territory | region | downloads | rank | country/territory | region | downloads |
| --- | --- | --- | --- | --- | --- | --- | --- |
| 1 | United States | North America | 392,985 | 16 | Switzerland | Europe | 13,728 |
| 2 | Germany | Europe | 179,660 | 17 | Malaysia | Asia | 13,184 |
| 3 | China | Asia | 151,489 | 18 | Austria | Europe | 12,987 |
| 4 | United Kingdom | Europe | 55,939 | 19 | Spain | Europe | 11,010 |
| 5 | Canada | North America | 38,511 | 20 | Hong Kong | Asia | 9987 |
| 6 | Australia | Australia | 38,486 | 21 | Mexico | Latin America | 9191 |
| 7 | Japan | Asia | 33,779 | 22 | Sweden | Europe | 9028 |

| 8 | India | Asia | 29,973 | 23 | Turkey | Asia | 8872 |
|---|---|---|---|---|---|---|---|
| 9 | Iran | Asia | 27,680 | 24 | Poland | Europe | 7416 |
| 10 | France | Europe | 27,230 | 25 | New Zealand | Australia | 6710 |
| 11 | South Korea | Asia | 22,047 | 26 | Belgium | Europe | 6430 |
| 12 | Taiwan | Asia | 20,928 | 27 | Denmark | Europe | 6089 |
| 13 | Netherlands | Europe | 19,841 | 28 | Portugal | Europe | 6017 |
| 14 | Brazil | Latin America | 17,880 | 29 | South Africa | Africa | 5669 |
| 15 | Italy | Europe | 13,796 | 30 | Thailand | Asia | 5211 |

We display the total downloads from different countries on a world map, as Fig. 2 shows. The node size reflects the downloads of the country. As can be seen, the United States, Germany and Mainland China are the top three biggest nodes. Most European countries have more than 1000 downloads, which makes the European part of the map very dense. And hence, we zoom in this region and make a partial enlarged view. Most South American countries and African countries have few downloads except Brazil and South Africa.

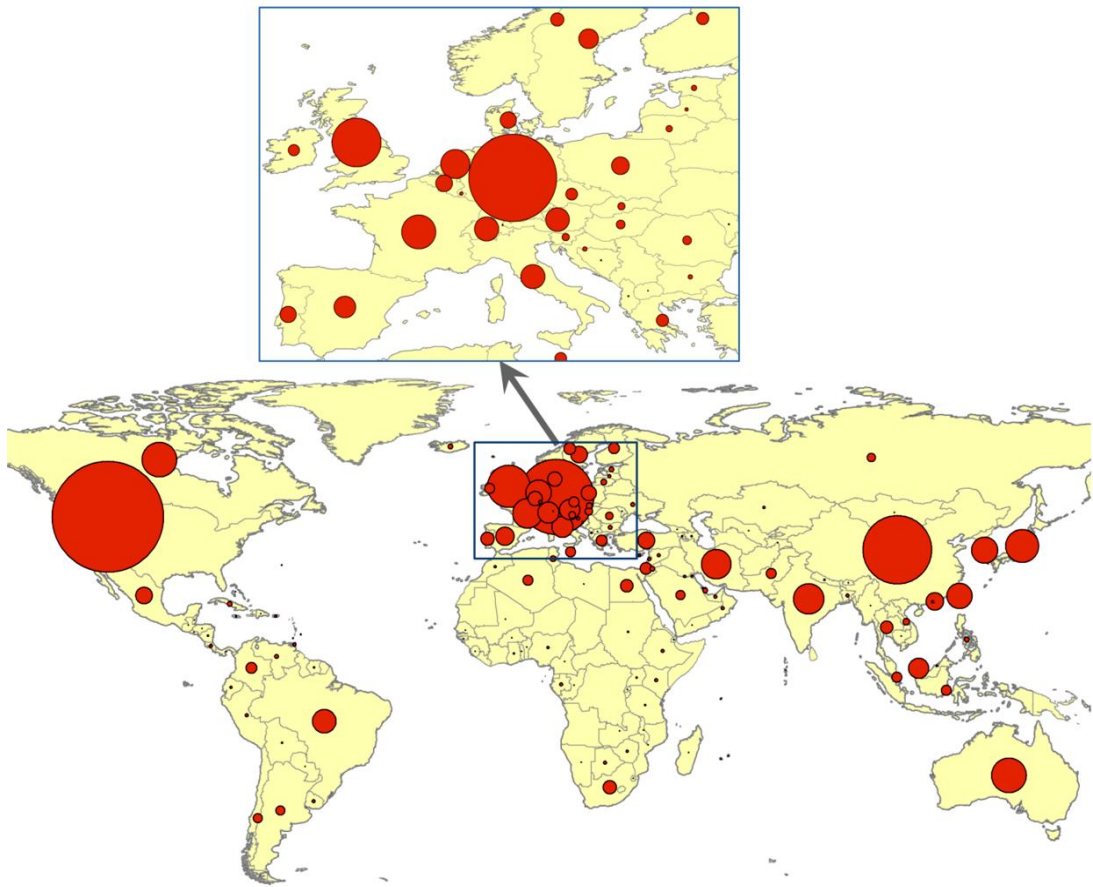

**Fig. 2** The world downloads map

Table 3 shows the total downloads of different regions. North America (including the United States and Canada) has a total of 431,500 downloads, accounting for 33.65% of the 1,128,378 downloads worldwide. Europe has 402,184 downloads and accounts for 31.36%. Asia has downloaded 346,812 items totally, accounting for 27.04% of all. And for Australia,

the number is 45,196 and 3.52%.

**Table 3** Total downloads of different regions during the research period

| region | downloads | percentage |
|---|---|---|
| North America | 431,500 | 33.65% |
| Europe | 402,184 | 31.36% |
| Asia | 346,812 | 27.04% |
| Australia | 45,196 | 3.52% |
| Others | 56,686 | 4.42% |
| Worldwide total | 1,282,378 | 100% |

**3.2 World downloads in every minute: peak and plateau**

Fig. 3(a) displays the downloads on 2 weekdays (April 11 and April 13) and 2 weekends (April 14 and April 15) in minutes. For both the 2 weekdays and the 2 weekends, the downloading curve is similar to a large extent.

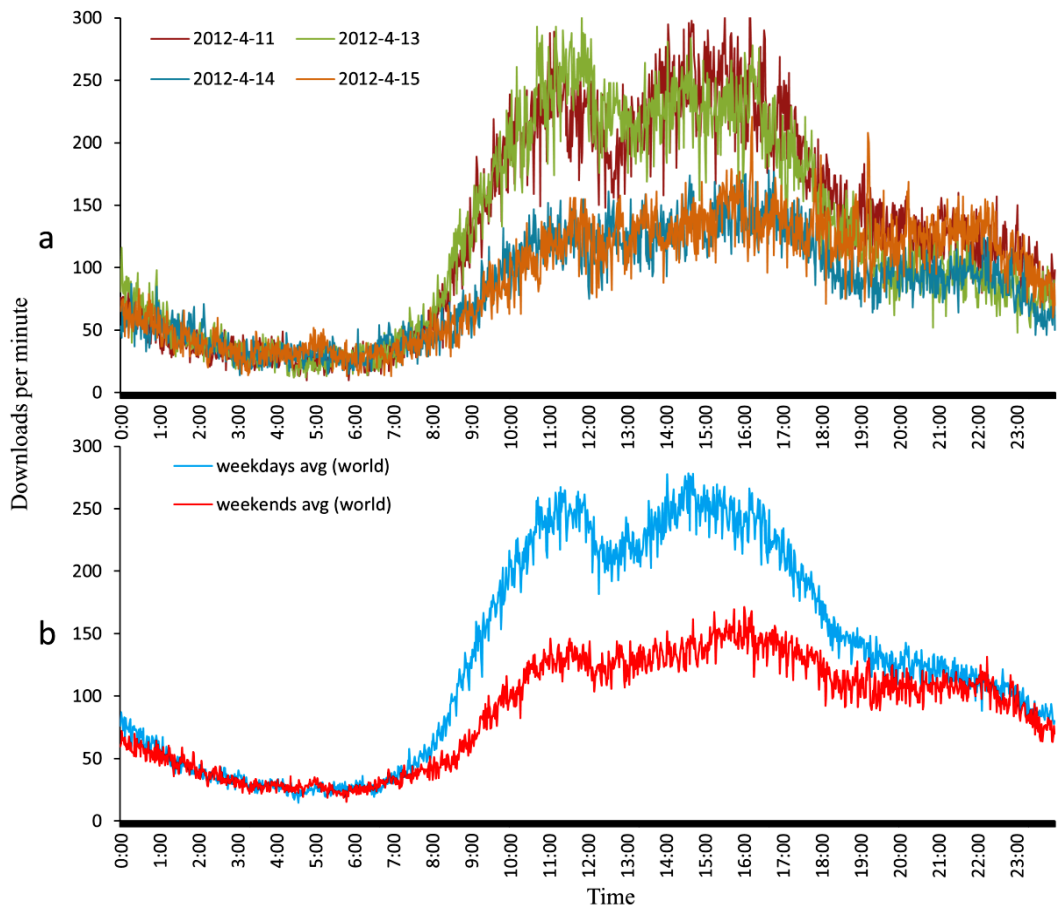

**Fig. 3(a-b)** Downloads of the world in every minute (local time)

Because the curves show very similar rhythms as Fig. 3(a) displays, we average the downloads on 4 weekdays (April 10, April 11, April 12, April 13)and 4 weekends (April 14, April 15, April 21, April 22) separately in Fig. 3(b). The blue curve shows weekday downloads and the red one represents weekend downloads.

Between the two peaks in the weekday downloading curve, there is a little valley around 12:00 to 13:00. However, for the weekends, the number of downloads keeps rising from about 07:00, and reaches a plateau rather than a summit. Then the weekend curve falls down to another relatively lower plateau at about 18:00, and at about 22:00 follows a rapid drop.

Fig. 4 displays the accumulated downloads of the world in every minute. Fig. 4(a) shows the downloads of the 8 days during our study period. Basically, the curves of 4 weekdays are very similar, so as the 4 weekends. Similarly, we calculate the average of the weekday and weekend downloads, and displays the curves in Fig. 4(b).

As is shown, before 9:00, the accumulated curves of weekdays and weekends are almost fully coincident. However, after 9:00, these two curves start to separate from each other. From the morning to the afternoon, the differentiation of the two curves becomes larger, and reaches the peak at around 18:00. And then, the gap lasts relatively constant until the end of the day.

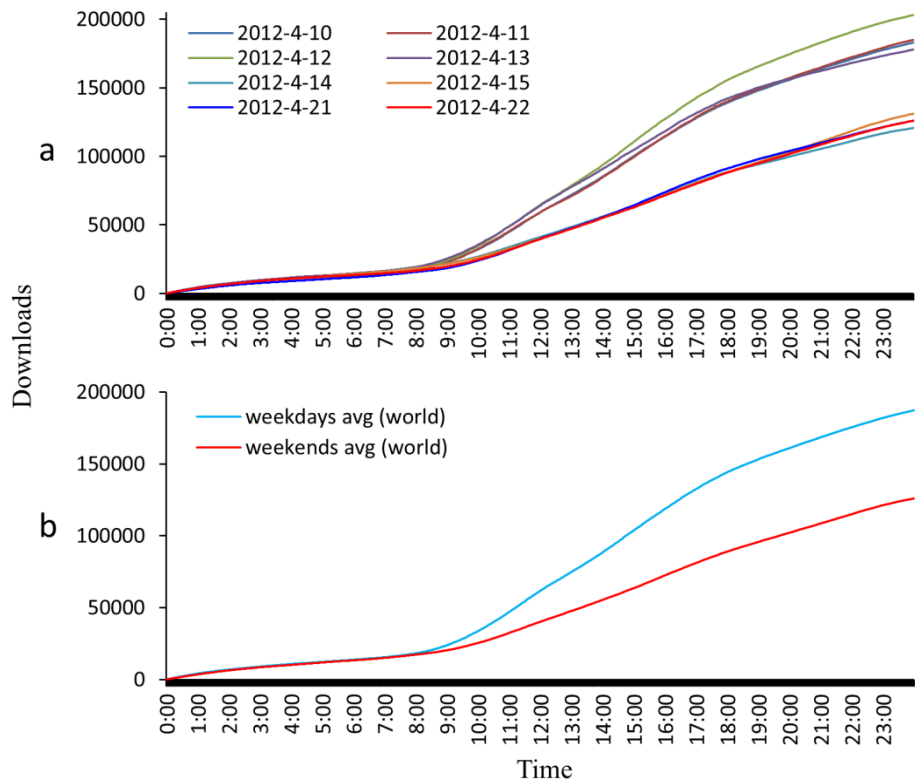

**Fig. 4(a-b)** Accumulated downloads of the world (local time)

**3.3 Country/territory downloading curves**

For each representative country/territory listed in Table 2, we analyze the downloading curves in detail.

***American Continent and Australian countries***

For weekdays, the number of downloads from the US keeps rising from about 08:00 and peaks around 16:00 in the afternoon. The continual rise of the US curve from 8:00 to the peak 16:00 means there is no fixed lunch time. And a large number of scientists work late and continue into the early morning hours, as Fig. 5(a) shows. On weekdays in Canada, from 8:00 in the morning, the downloading curve gradually climbs to the first peak at about 12:00. Then, after a slight decline, the second summit comes around 15:00, but followed by a rapid downward trend from 17:00 to 18:30. After 18:30, the downloading curve keeps a low trend, as shown in Fig. 5(b).

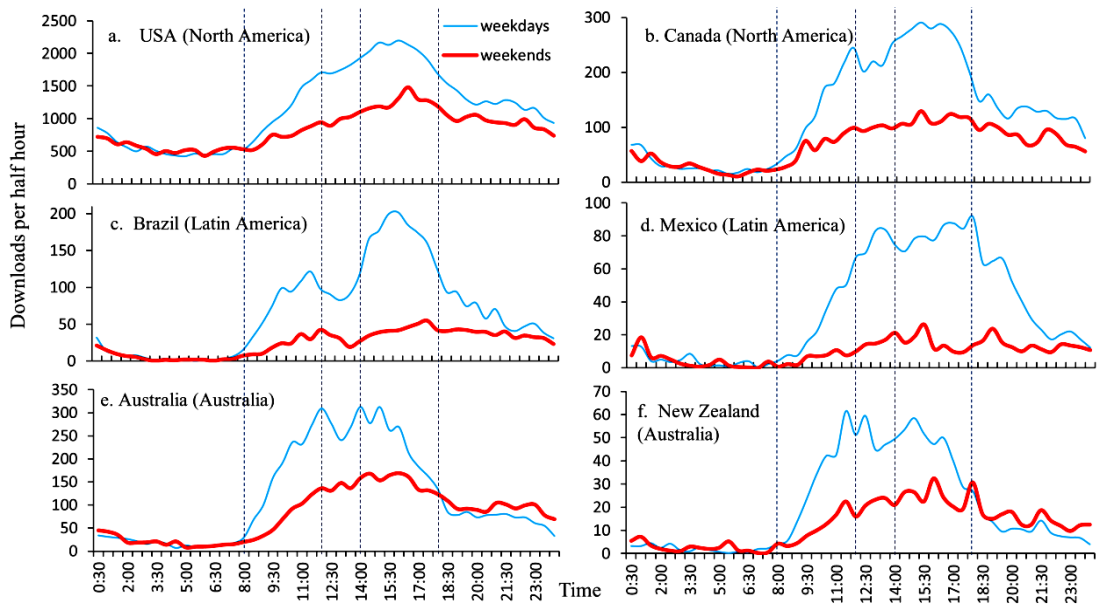

**Fig. 5(a-f)** Downloads of North American and Latin American countries (local time)

Brazil and Mexico are the two most downloading countries in Latin America. In weekdays' afternoons, Brazilian scientists work much more than in the morning. For Mexico, on weekdays, the downloading curve climbs to a summit at around 13:00, followed by a relatively long fluctuation, and starts a quick downtrend at 18:00. On weekends, both of the two flat red curves in Fig. 5(c) and Fig. 5(d) indicate that scientists are not so busy with their work on weekends.

The Australian curve and New Zealand curve show a similar trend on both weekdays and weekends. As shown in Fig. 5(e) and Fig. 5(f). On weekdays, the curve starts to rise from around 8:00 to the summit around 12:00, and then, after a slight drop before 13:00, it continues to climb to a new plateau. As for the weekend downloading curve, both countries show a relatively flat curve. Moreover, at weekend night, the number of downloads even exceeds the number at weekday night.

*Asian countries/*territories

In China, we can see that a typical Chinese scientist's day is divided into three working periods by 2 troughs, which are lunch and dinner time. Chinese scientists tend to put work aside to have a rest during lunch and dinner time, and thus shaping two obvious valleys around 12:00 and 18:00. For the 18:00 valley, another explanation is that Chinese scientists start the third working period of one day after dinner. Although Chinese scientists don't work late at night, they work hard on the weekends as weekdays. As Fig. 6(a) shows.

For Japan, the shape of the weekday downloading curve seems like the Fujiyama."The volcanic crater" is the lunch time and lunch break, however, the curve climes to the summit around 15:00 rapidly. For weekends, the Japan downloading curve shows more gentle, and the number of downloads on the "small hills" is just over 100 every half hour. It seems that Japanese scientists always enjoy their weekends. As shown in Fig. 6(b).

In the case of India (Fig. 6(c)), downloads also reach two summits on weekdays, but it is obvious that the peak at noon is much higher than the one in the afternoon (like China). The curve begins to fall at lunch time at around 13:00, bottoms out at 14:00, and reaches the second summit at 15:00. On weekends, the curve presents a similar but more moderate trend.

Fig. 6(d) shows the Iranian downloading curves on weekdays and weekends. Moreover, different from most other countries, the curves have only one peak, which is around 11:00, on both weekdays and weekends. According to our result, Iranian scientists tend to download more papers on "weekends" than "weekdays". In fact, Friday is legal weekend in Iran. Thursday afternoon is also often regarded as weekend by most Iranians. Saturday and Sunday are weekdays by contrast.

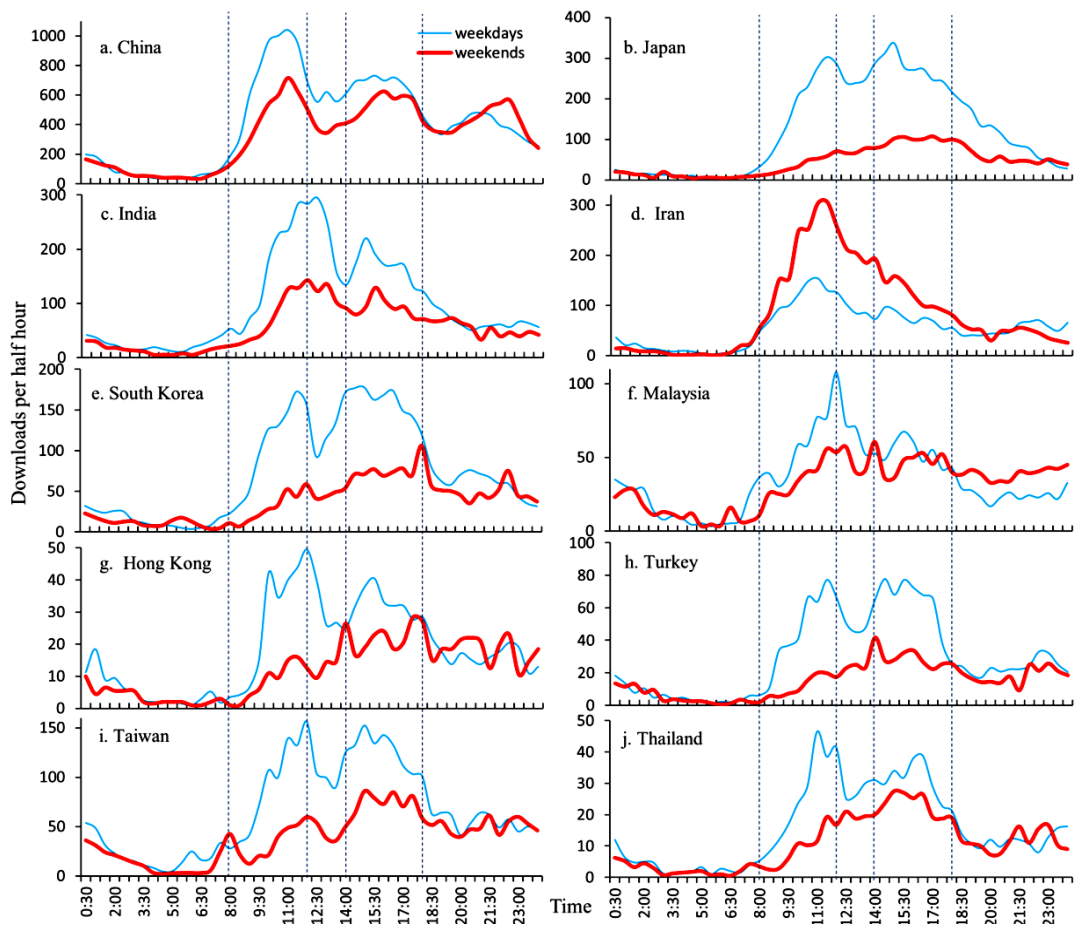

**Fig. 6(a-j)** Downloads of Asian countries/territories (local time)

The South Korean curves are quite consistent with Taiwanese and Turkish curves on weekdays. As Fig. 6(e), Fig. 6(i), and Fig. 6(j) show. There is a relatively deep groove around 11:30 to 13:30, which is the lunch time and lunch break. In addition, the morning "mountains" have the same height with the afternoon ones. On the weekdays, Hong Kong (Fig. 6(h)) also has a roughly similar curve with Mainland China. However, there aren't obvious regularities at weekends for the four countries and territories.

In Malaysia, unlike its famous building Kuala Lumpur City Centre (KLCC), the curve shows the single spire around 12:00 on weekdays. The noon break is from around 13:00 to 15:00, and the noon bottom is at 14:00. As is shown in Fig. 6(f). Malaysian scientists have busy weekends, especially busy weekend nights. The scientists in Thailand, another Southeast Asian country, also work hard in weekend afternoons.

*European and African countries*

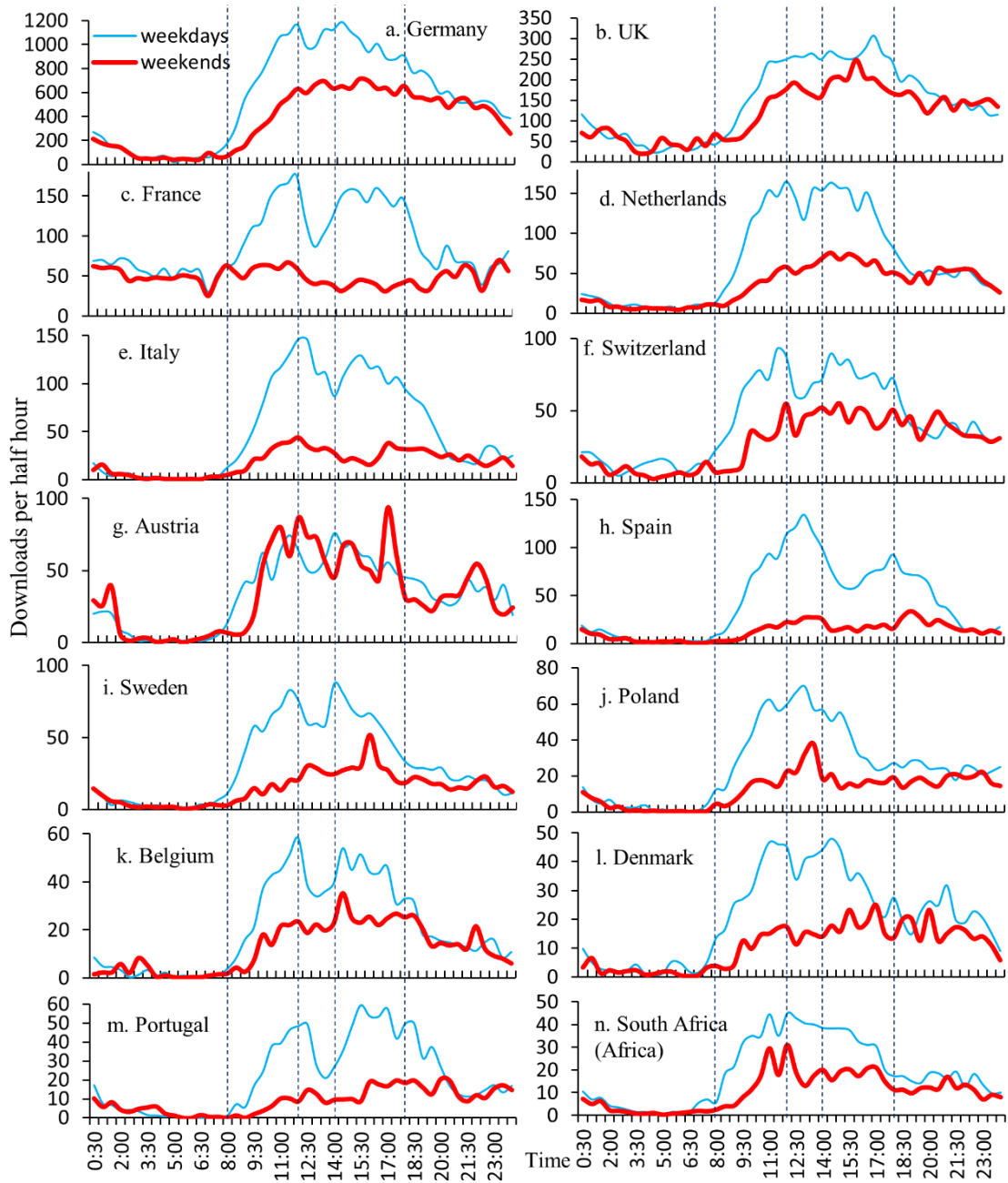

**Fig. 7(a-n)** Downloads of European and African countries (local time)

As Fig. 7 (a) shows, in Germany, on weekdays, the first climax comes around 11:00 in the morning. And then, after a slight decline around 12:00, the lunch time, the second summit comes around 13:00, followed by a fluctuant downward trend afterwards. For weekends, in spite of the relatively lower downloads compared with weekdays, the downloading peak and plateau last till night.

Like USA, there is no fixed lunch time for The United Kingdom, see Fig. 7 (b). From 10:00 to 18:00, the number of downloads keeps on a plateau. The gap between weekdays and weekends is very small, which means busy weekends for British scientists. Moreover, at weekend nights, the downloading number even exceeds the number at weekday nights.

It seems that some European countries have the same working patterns during the weekdays. The weekdays' curves of France, Italy, Switzerland, Belgium and Portugal look very similar. The only distinct difference is that Italy's lunch break is a little longer and later than the others, and the afternoon summit (at 15:30) is a little higher than the morning peak (at 12:30) in Portugal. As for weekends, the curve of France keeps a fluctuation within a narrow range of 50 downloads on a whole day, as shown in Fig. 7(c).

Like Germany, lunch break in Netherlands and Denmark is short. Dutch scientists keep the similar and well working condition in mornings and afternoons. Unlike it, Denmark has the third working period after dinner. As Fig. 7(d) and Fig. 7(i) show.

Spain also has two summits on weekdays, and the morning mountain is much higher than the afternoon one. Another point which is worth mentioning is that Spain's lunch break comes much later than other countries. The noon break takes a long time, at around 15:30. As shown in Fig. 7(h).

The weekend curve for some countries, particularly in southern Europe, is very low and flat, including Italy, Spain and Portugal. In spite of hard work on weekdays, the scientists in these countries still regard the weekend as rest time. However, for Austria, a central European country, it's difficult to distinguish the weekends curve from the weekday curve, as Fig. 7 (g) shows.

As is shown in Fig. 2, except South Africa, Egypt and Algeria, there are only few downloads in African continent. South Africa is also the only African country listed in the top 30 countries. Fig. 7 (n) displays the downloading curves of South Africa. The two curves show very coincident rhythm. The lunch break is at around 11:30, and is very short.

*Special downloading phenomena*

Special downloading phenomena are different from the error data mentioned in the data processing section. The error data illustrate the extremely abnormal downloading burst from minute to minute for one IP address, but for the special downloading phenomena mentioned here, the downloading is still normal for IP addresses, but the downloading curve is quite different.

For most countries, all the downloading curves are separately similar, whether on weekdays or weekends. However, below are some abnormalities.

**(1) Korean election**

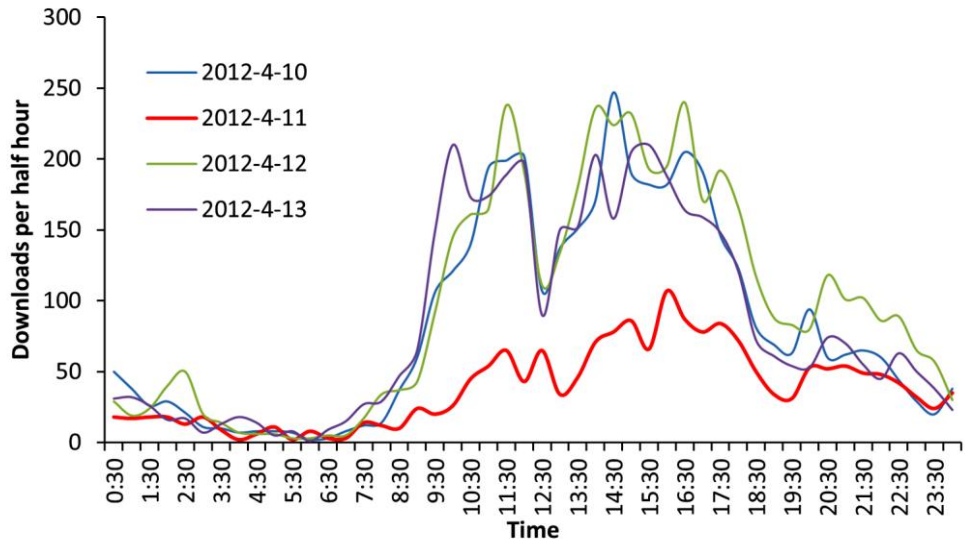

**Fig. 8** Abnormal downloads of South Korea (local time)

As is shown in Fig. 8, the downloading curves on 3 weekdays of South Korea are relatively coincident. Nevertheless, the curve of April 11 is very different from the others. This is partly affected by the elections for the members of unicameral National Assembly, which is held every 4 years.

**(2) Iranian weekends**

Specificities are also found in the data of Iran. Fig. 9 displays the downloads of 8 days. We find that the curve of April 13 (Friday) is subaverage obviously, which is consistent with the legal weekend day in Iran. The downloads of April 22 (Sunday) is much higher than the other days.

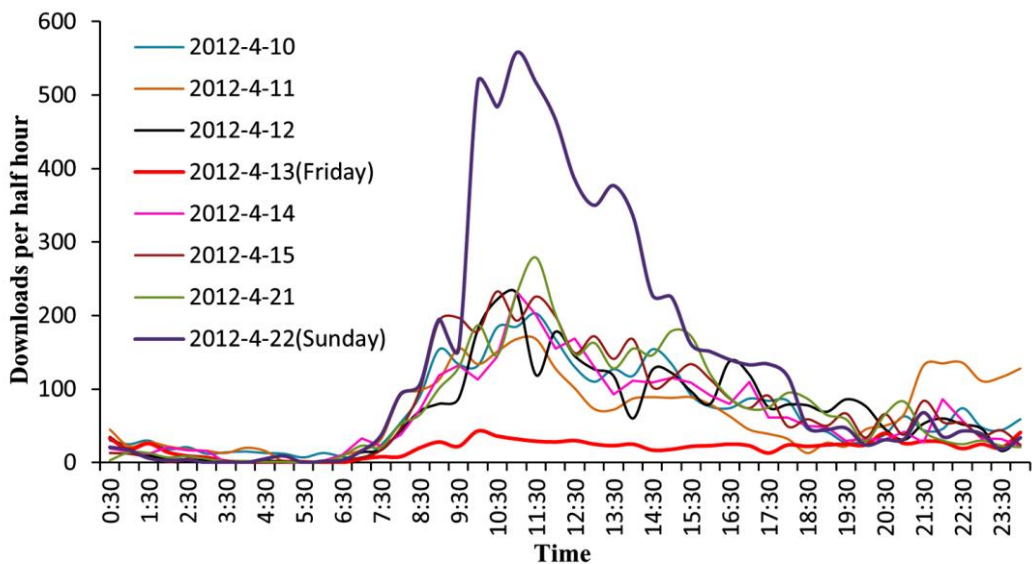

**Fig. 9** Daily downloads of Iran (local time)

## 4. Discussion

Different working time patterns in different countries are revealed in our study. In the United States and France, overnight work is more prevalent among scientists; while in the United Kingdom and Mainland China, scientists usually work almost as hard on the weekends as on the weekdays. Moreover, from the analysis results of South Korea and Iran, we gain a preliminary understanding of the role of holidays on scientists' working timetable. Furthermore, other social factors, e.g. culture, politics and religion, partly have an impact on scientists' research activities.

Nevertheless, despite the different working habits in different countries, a quite common conclusion in most countries is that scientists all over the world today are working overtime. Both their mental and physical health is negatively affected by overwork. Meanwhile, the ambiguity of the boundary between home and office causes severe work-family conflicts, which means the influence does not limit to scientists themselves.

Scientists may feel engaged and satisfied with their research work (Russo, 2012), however, working too much is still a serious and heavy issue, which warns us to reconsider the work-life balance.

## Acknowledgements

The work was supported by the project of "Social Science Foundation of China"(10CZX011) and the project of "Fundamental Research Funds for the Central Universities" (DUT12RW309).